\documentclass[10pt]{article}

\usepackage{fullpage}
\usepackage{setspace}
\usepackage{parskip}
\usepackage{titlesec}
\usepackage{placeins}
\usepackage{xcolor}
\usepackage{float}
\usepackage{lineno}
\usepackage{breqn}
\usepackage[toc,page]{appendix}
\usepackage{physics}
\usepackage[numbers,sort&compress]{natbib}
\newcommand{\keywords}[1]{\textbf{Keywords:} #1}

\usepackage{atbegshi}% http://ctan.org/pkg/atbegshi
\AtBeginDocument{\AtBeginShipoutNext{\AtBeginShipoutDiscard}}

\PassOptionsToPackage{hyphens}{url}
\usepackage[colorlinks = true,
            linkcolor = blue,
            urlcolor  = blue,
            citecolor = blue,
            anchorcolor = blue]{hyperref}
\usepackage{etoolbox}
\makeatletter
\patchcmd\@combinedblfloats{\box\@outputbox}{\unvbox\@outputbox}{}{%
}%
\makeatother

\renewenvironment{abstract}
  {{\bfseries\noindent{\large\abstractname}\par\nobreak}}

\titlespacing{\section}{0pt}{*3}{*1}
\titlespacing{\subsection}{0pt}{*2}{*0.5}
\titlespacing{\subsubsection}{0pt}{*1.5}{0pt}

\usepackage{authblk}

\usepackage{graphicx}
\usepackage[space]{grffile}
\usepackage{latexsym}
\usepackage{textcomp}
\usepackage{longtable}
\usepackage{tabulary}
\usepackage{booktabs,array,multirow}
\usepackage{amsfonts,amsmath,amssymb}
\providecommand\citet{\cite}
\providecommand\citep{\cite}

\newif\iflatexml\latexmlfalse
\DeclareGraphicsExtensions{.pdf,.PDF,.png,.PNG,.jpg,.JPG,.jpeg,.JPEG}

\usepackage[utf8]{inputenc}
\usepackage[english]{babel}

\begin{document}
\newcommand{\Sq}[1]{ \frac{ #1 }{ \sqrt{2} } }
%Title of paper
\title{Time resolved quantum tomography in molecular spectroscopy  by the Maximal Entropy Approach}

\author[1]{Varun Makhija
\textsuperscript{\ddag}}
\thanks{vmakhija@umw.edu}
\affil[1]{Department of Chemistry and Physics, University of Mary Washington, Fredericksburg, Virginia 22401, USA }

\author[2]{Rishabh Gupta\textsuperscript{\ddag}}
% \thanks{These authors contributed equally to this work.}}
\affil[2]{Department of Chemistry, Purdue University, West Lafayette, Indiana, 47907, USA}

\author[3]{Simon Neville}
\author[3,4]{Micheal Schuurman}
\affil[3]{National Research Council Canada, 100 Sussex Drive, Ottawa, Ontario, K1A 0R6, Canada\\}
\affil[4]{Department of Chemistry, University of Ottawa, Ottawa, Ontario, K1N 6N5, Canada\\}

\author[5]{Joseph Francisco}
\affil[5]{Department of Earth and Environmental Science and Department of Chemistry, University of Pennsylvania, Philadelphia, PA, USA}

\author[6]{Sabre Kais}
\thanks{kais@purdue.edu}
\affil[6]{Department of Chemistry, Department of Electrical and Computer Engineering, and Purdue Quantum Science and Engineering Institute,  Purdue University, West Lafayette, Indiana, 47907, USA}

\vspace{-1em}
  \date{\today}
\begingroup
\let\center\flushleft
\let\endcenter\endflushleft
\maketitle
\endgroup

\renewcommand{\thefootnote}{\ddag}
\footnotetext{These authors contributed equally to this work.}
\renewcommand{\thefootnote}{\arabic{footnote}}

\begin{abstract}
Attosecond science offers unprecedented precision in probing the initial moments of chemical reactions, revealing the dynamics of molecular electrons that shape reaction pathways. A fundamental question emerges: what role, if any, do quantum coherences between molecular electron states play in photochemical reactions? Answering this question necessitates quantum tomography—the determination of the electronic density matrix from experimental data, where the off-diagonal elements represent these coherences. The Maximal Entropy (MaxEnt) based Quantum State Tomography (QST) approach offers unique advantages in studying molecular dynamics, particularly with partial tomographic data. Here, we explore the application of MaxEnt-based QST on photoexcited ammonia, necessitating the operator form of observables specific to the performed measurements. We present two methodologies for constructing these operators: one leveraging Molecular Angular Distribution Moments (MADMs) which accurately capture the orientation-dependent vibronic dynamics of molecules; and another utilizing Angular Momentum Coherence Operators to construct measurement operators for the full rovibronic density matrix in the symmetric top basis. A key revelation of our study is the direct link between Lagrange multipliers in the MaxEnt formalism and the unique set of MADMs. Additionally, we visualize the electron density within the molecular frame, demonstrating charge migration across the molecule. Furthermore, we achieve a groundbreaking milestone by constructing, for the first time, the entanglement entropy of the electronic subsystem—a metric that was previously inaccessible. The entropy vividly reveals and quantifies the effects of coupling between the excited electron and nuclear degrees of freedom. Consequently, our findings open new avenues for research in ultrafast molecular spectroscopy within the broader domain of  quantum information science, offering profound implications for the study of molecular systems under excitation using quantum tomographic schemes.
\end{abstract}

\section{Introduction}\label{sec1}

The field of Molecular Reaction Dynamics seeks to understand the detailed evolution of chemical reactions at the molecular level ~\cite{zewail2000,levine2009m}. The development of attosecond physics allows for experimental access to such molecular details during the initial moments of photochemical processes with unprecedented time resolution, with the promise of exerting control on such reactions at these timescales~\cite{li2022, chang2020, zinchenko2021, ruberti2021, huppert2016, calegari2014, dixit2012,matselyukh2022,ranitovic2014,suominen2014, santra2014, despre2018,sandor2018, simmermacher2019,hermann2020,carrascosa2021,folorunso2021,dey2022,plunkett2022,matselyukh2022}. As suggested by experimental and theoretical studies, these initial dynamics are driven by quantum mechanical coherences between molecular electronic states, and coupling of the electronic motion to the nuclear degrees of freedom~\cite{ruberti2021,matselyukh2022,blanchet1999,underwood2008,kowalewski2015,arnold2017,neville2016,dey2022}. The direct measurement of such coherences necessitates time resolved quantum tomography on these timescales~\cite{zhang2021, jiang2024}, which then also facilitates imaging the resulting dynamics~\cite{gregory2022, morrigan2023}. Thus far, this has only been successfully achieved in the electronic ground state \cite{dunn1995,skovsen2003,mouritzen2006,hasegawa2008,zhang2021} and the limited case of an excited two-electronic-state system~\cite{morrigan2023}. Furthermore, the latter demonstration relied on prior knowledge of symmetry selection rules into the excited states, and applies only to a specific measurement technique. In quantum information science (QIS), quantum tomography has been a subject of intensive study, as quantum computation necessitates the robust determination of quantum states prepared in the lab~\cite{dariano2003,gupta2021,gupta2021convergence,gupta2022, gupta_hamiltonian}. Within QIS, Quantum State Tomography (QST) endeavors to derive statistical inferences regarding an unknown quantum state within a system. This process hinges on extracting insights from the expectation values of a comprehensive set of observables~\cite{nielson,kais,photonic, qubits, cramer}. Traditionally, these observables correspond to the mean measurement values of an informationally complete set of Hermitian operators in order to uniquely reconstruct an unknown quantum state. However, the conventional approach to QST faces significant challenges when applied to modern large-scale quantum platforms ~\cite{10-qubit}. This challenge stems from the exponential increase in the number of measurements required for complete state reconstruction relative to the size of the system. Consequently, performing QST for even moderately sized systems represents a substantial difficulty. In addition to the formidable challenge of exponential scaling in analyzing complex systems, the quest for accurate estimation of quantum states encounters yet another hurdle due to the pervasive noise inherent in today's Noisy Intermediate-Scale Quantum (NISQ) devices ~\cite{preskill2018quantum}. Consequently, the fidelity of available measurements is compromised, with some measurements even proving inaccessible. To circumvent these limitations, researchers have proposed diverse approaches diverging from traditional QST. These innovative methods include Neural Network Tomography ~\cite{Torlai2018, Carrasquilla2019, Palmieri2020, xin2019local}, Matrix Product State Tomography ~\cite{cramer}, Quantum Overlapping Tomography ~\cite{overlap}, and Shadow Tomography ~\cite{aaronson, Huang2020}.
Recognizing the noisy reality of quantum systems and the challenge of obtaining high-fidelity measurements across all parameters, efforts have emerged to conduct QST based on incomplete data. Methodologies such as Maximum Likelihood Estimation (MLE) ~\cite{hradil, teo2011quantum, teo2012incomplete, blume2010hedged, smolin2012efficient, baumgratz2013scalable}, Bayesian Mean Estimation (BME) ~\cite{blume2010optimal, huszar2012adaptive, lukens2020practical, lukens2020bayesian}, and Maximal Entropy Formalism-based QST ~\cite{gupta2021,gupta2021convergence,gupta2022, D2CS00203E} have been explored in pursuit of more robust estimation methodologies. \\ 
To the best of our knowledge, tomographic methods developed in QIS have not yet been applied to dynamics in a photoexcited molecule. Potentially, these methods could generalize to several techniques of ultrafast measurement. Here, we demonstrate the application of the Maximal Entropy Approach (MaxEnt) to quantum state tomography for the photoexcited ammonia molecule, the same two-electronic-state system studied in~\cite{morrigan2023}. This represents four important advances in ultrafast molecular spectroscopy: (i) MaxEnt does not require prior knowledge of symmetry selection rules in the excited state; (ii) To apply MaxEnt, we determine an operator form of the observable used in the experiment in~\cite{morrigan2023}, which can potentially be applied to other ultrafast measurement techniques; (iii) MaxEnt can be used in cases of `incomplete' information~\cite{hou2022}, when a unique tomography is not possible, to determine a compatible quantum state, allowing its extension to cases with more degrees of freedom; (iv) MaxEnt allows construction of the time evolving entropy of the electronic subsystem directly revealing and quantifying the correlation of electronic motion with nuclear degrees of freedom, and tracking the dynamics of quantum information in the molecule \cite{zhang2022,zhang2024}. These advances set the foundation for the generalization of quantum tomography in ultrafast molecular reaction dynamics. Moreover, the results presented here provide the first physical interpretation for the parameters in the MaxEnt approach to quantum tomography. This demonstrates the possibility of feedback between time-resolved quantum dynamics and QIS that can be generated by future applications of our methods in ultrafast science. 

\section{Maximal Entropy Approach to Quantum Tomography} \label{entropy}
The MaxEnt formalism is a powerful and elegant framework in statistical physics and information theory that allows us to derive the most unbiased probability distribution, given a set of constraints~\cite{raphy,alhassid}. At its core, MaxEnt seeks to maximize the entropy of a system while honoring the constraints that capture our prior knowledge about the system. This formalism provides a robust and principled approach to infer the probability distribution of a classical system when we have incomplete information or limited data. \\
One of the key insights of MaxEnt lies in its seamless integration with the method of Lagrange multipliers ($\lambda_k \in \mathbf{C}^2$)~\cite{Jaynes,Jaynes2,jaynes3,wich}, which introduces a set of parameters that enable us to craft probability distributions that satisfy specific constraints. When applied to quantum systems, MaxEnt leads us to a remarkable expression for the density operator ($\hat{\rho}_{ME}$) of an unknown quantum state as it seeks to maximize the von Neumann entropy of the system. This density operator encapsulates the essence of MaxEnt's power—it yields a density matrix that is both least biased as well as consistent with our preconceived knowledge, as represented by the known expectation values of certain operators $\hat{f}_k$. 
\begin{equation}
\hat{\rho}_{ME}(\Vec{\lambda},\hat{f}) = \frac{1}{Z(\lambda_{1},\ldots,\lambda_{k})}\exp\{-\sum_{k}\lambda_{k}\hat{f}_{k}\} \label{rho2}
\end{equation}
The expression for $\hat{\rho}_{ME}$, as shown in Eq. (\ref{rho2}), elegantly combines the Lagrange multipliers $\lambda_k$ and the known operators $\hat{f}_k$. The exponential term captures the interplay between these parameters, allowing us to precisely balance the desire for maximizing entropy with the constraints imposed by our prior information. The normalization factor $Z(\lambda_{1},\ldots,\lambda_{k})=\Tr\{\exp\{-\sum_{k}\lambda_{k}\hat{f}_{k}\}\}$ guarantees that the resulting density operator represents a valid density matrix. In essence, MaxEnt empowers us to strike a harmonious balance between complete neutrality, represented by maximum entropy, and the real-world constraints that govern our understanding of a system. \\
Within the context of quantum tomography, reconstruction of an unknown quantum state relies on experimental measurements of Hermitian operators, typically defined as Positive-Operator-Valued Measures (POVMs). These measurements provide essential information about the quantum system, enabling the inference of its characteristics through reconstruction techniques. The MaxEnt formalism has proven effective in reconstructing quantum states prepared on quantum devices across various \textit{N}-qubit systems. In these cases, constraints are imposed by the expectation values of operators representing probabilities and coherences. This approach has also been implemented on prototype quantum devices provided by IBM that also corroborates the MaxEnt method for quantum tomography~\cite{gupta2021,gupta2021convergence,gupta2022}. \\
It is important to note that, unlike typical QST procedures for quantum devices, the operators used in MaxEnt-based QST for molecular dynamics do not need to be POVMs. For molecular dynamics a time and orientation angle-dependent Lab Frame Density Matrix (LFDM), $\rho_{ij}(\mathbf{\Omega},t)$, with $\mathbf{\Omega}$ representing the orientation angles, can be defined in a basis of electronic-vibrational (vibronic) states $\ket{i}$, $\ket{j}$~\cite{gregory2022}.  Application of MaxEnt to determine this LFDM requires the definition of operators $\hat{O}_i(\mathbf{\Omega})$, representing observables, such that 
\begin{equation}
\langle \hat{O}_i \rangle (t) = \Tr\left\{\hat{\rho}(\mathbf{\Omega},t)\hat{O}_i(\mathbf{\Omega})\right\},
\label{eq:traceM}
\end{equation}
where the trace is over all molecular coordinates including orientation angles, which we refer to as the `full trace' here in. If a number of time resolved measurements $\langle O_i \rangle (t)$ are made, the maximal entropy density matrix satisfying Eq. (\ref{eq:traceM}) may be represented as~\cite{gupta2021,hou2022}
\begin{equation}
\rho(\mathbf{\Omega},t) = \frac{e^{-\sum_{i}\lambda_i(t)\hat{O}_i(\mathbf{\Omega})}}{Z}
\label{eq:maxent}
\end{equation}
where $Z=\Tr\{e^{-\sum_{i}\lambda_i(t)\hat{O}_i(\mathbf{\Omega})}\}$, with $\Tr$ being the full trace, and the $\lambda_i(t)$ are Lagrange multipliers determined by fitting to the experimental data. This approach has properties that make it particularly suitable to the molecular case. Specifically, it is applicable to mixed states and it can be used to find a state compatible with Eq. (\ref{eq:traceM}) and (\ref{eq:maxent}) in cases where complete tomographic information is not available~\cite{hou2022}. The application of MaxEnt to molecular dynamics requires knowing the form of $\hat{O}_i(\mathbf{\Omega})$ for the particular measurement performed. This is constructed in Section \ref{operator_form} below. 

\section{The Operator form of the observable} \label{operator_form}
We discuss two possible ways to construct operators for time resolved photoelectron angular distributions measured from a resonantly excited molecule~\cite{underwood2008}. One uses the previously studied Molecular Angular Distribution Moments (MADMs)~\cite{makhija2020,gregory2022,morrigan2023} to explicitly select the $\ket{\mathbf{\Omega}}$ - diagonal elements of the vibronic density matrix, the LFDM, probed by the experiment~\cite{gregory2022,underwood2008}. This representation has previously been used to determine molecular frame quantities from lab frame measurements~\cite{makhija2016,makhija2020,lam2020, sandor2018, sandor2019, marceau2017, gregory2021, lam2022}. For a recent review see~\cite{hockett2023}. It may be considered an irreducible representation of the measurement operator, which provides the necessary information to fully determine only the molecular frame quantities. Alternatively, a measurement operator which comprises the full rovibronic density matrix in the symmetric top basis can be constructed using the Angular Momentum Coherence Operators introduced in~\cite{gregory2021}. This representation contains redundant information in molecular frame, but new lab frame information. Only the LFDM has been successfully constructed from experimental data~\cite{morrigan2023}, but both options can potentially be addressed by MaxEnt~\cite{gupta2021,gupta2021convergence,gupta2022}. While we develop both options here, we focus on reconstructing the LFDM leaving the rovibronic problem for future study.    
\subsection{Option One: The LFDM}
The LFDM, determining the subspace diagonal in the Euler angle basis of molecular orientations $\ket{\mathbf{\Omega}}$, is~\cite{gregory2022}
\begin{equation}
\hat{\rho}(\mathbf{\Omega},t) = \sum_{ij}\int{d\mathbf{\Omega}}\sum_{KQS}A^{K}_{QS}(i,j;t)D^{K*}_{QS}(\mathbf{\Omega})\ket{i}\ket{\mathbf{\Omega}}\bra{\mathbf{\Omega}}\bra{j}.
\label{eq:LFDMop}
\end{equation}
Here, the $A^K_{QS}(i,j;t)$ are the previously studied MADMS~\cite{gregory2021} associated with vibronic states $\ket{i}$ and $\ket{j}$ and $D^K_{QS}(\mathbf{\Omega})$ are elements of the Wigner D-matrix representation of the rotation operator. In this basis, we also define a measurement operator
\begin{equation}
\hat{O}_{LM}(\mathbf{\Omega};\epsilon)=\sum_{ij}\int{d\mathbf{\Omega}}\sum_{KQS}\frac{2K+1}{8\pi^2}C^{LM}_{KQS}(j,i;\epsilon)D^{K}_{QS}(\mathbf{\Omega})\ket{i}\ket{\mathbf{\Omega}}\bra{\mathbf{\Omega}}\bra{j}.
\label{eq:measop}
\end{equation}
The coefficients $C^{LM}_{KQS}(j,i;\epsilon)$ are determined by the nature of the probe process, $\epsilon$ being parameters associated with this process. For example, in the case of ejected photoelectrons from an excited molecule, $\epsilon$ is the kinetic energy of the ejected electron. In general, it is assumed that the scattered particle used as a probe has some anisotropy in the lab frame, allowing a multipole expansion of the detected signal in spherical harmonics $Y_{LM}(\theta_e,\phi_e)$, $\theta_e$ and $\phi_e$ being lab frame ejection angles. From this, we can determine the time-dependent measured quantity $\Tr\{\hat{\rho}(\mathbf{\Omega},t)\hat{O}_{LM}(\mathbf{\Omega};\epsilon)\}$. This can be achieved using the following well-known orthonormality relations: 
\begin{equation}
\int{d\mathbf{\Omega}D^{K*}_{QS}(\mathbf{\Omega})D^{K'}_{Q'S'}(\mathbf{\Omega})}=\frac{8\pi^2}{2K+1}\delta_{KK'}\delta_{QQ'}\delta_{SS'};  
\end{equation}
which, along with Eq.~\ref{eq:LFDMop} and Eq.~\ref{eq:measop}, lead to 
\begin{equation}
\left(\hat{\rho}(\mathbf{\Omega},t)\hat{O}_{LM}(\mathbf{\Omega};\epsilon)\right)_{ii}=\sum_{j}\sum_{KQS}C^{LM}_{KQS}(i,j;\epsilon)A^{K}_{Q,S}(i,j;t)
\end{equation}
for the diagonal elements of $\hat{\rho}(\mathbf{\Omega},t)\hat{O}_{LM}(\mathbf{\Omega};\epsilon)$. The resulting trace yields the desired measured quantity:
\begin{equation}
\beta_{LM}(\epsilon;t)=\sum_{ij}\sum_{KQS}C^{LM}_{KQS}(i,j;\epsilon)A^K_{QS}(i,j;t).
\label{eq:data}
\end{equation}
For the case of measurement by one-photon ionization the coefficients $C^{LM}_{KQS}(i,j;\epsilon)$ contain matrix elements of the dipole operator between the states $\ket{i}$, $\ket{j}$ and the continuum states used as a template for measurement~\cite{underwood2008}. These are determined by `complete' photoionization experiments, typically using high-resolution resonant spectroscopies~\cite{hockett20071,hockett20072}. We note that Eq.~\ref{eq:measop} does not represent the measurement of an equivalent quantity for a molecule fixed at orientation $\mathbf{\Omega}$ relative to the lab frame, but instead rigorously defines the operator associated with the lab frame measurement of the quantity in Eq.~\ref{eq:data}. In some cases, such as if only the $L = 0$, $M = 0$ term is considered (the ejection-angle integrated yield), these are equivalent. For one-photon ionization, a transformation matrix exists for symmetric molecules that provides the molecule-fixed $\beta_{LM}(\epsilon;\mathbf{\Omega})$ if the $C^{LM}_{KQS}(i,j;\epsilon)$ are known. Recent studies of one and few-photon ionization have shown that these time resolved measurements are also parameterized by Eq.~\ref{eq:data}, and that the $C^{LM}_{KQS}(i,j;\epsilon)$ can be determined for ionization of the electronic ground state through ionization of an excited rotational wavepacket~\cite{marceau2017, gregory2021,lam2022}. Similarly, ultrafast X-ray and electron diffraction measurements can also be parameterized by Eq.~\ref{eq:data} as shown in ~\cite{hegazy2023}, in which case the $C^{LM}_{KQS}(i,j;\epsilon)$ were also determined
in the ground state by diffraction from a rotational wavepacket. Eq.~\ref{eq:data} has also been extensively used to parameterize measurements of strong field ionization and dissociation ~\cite{makhija2016,makhija2020,lam2020, sandor2018, sandor2019, marceau2017, gregory2021, lam2022}. Thus, Eq.~\ref{eq:data} can serve as a general way to parameterize ultrafast measurements in terms of matrix elements of the operator Eq.~\ref{eq:LFDMop}, the two being directly related by the LFDM: $\beta_{LM}(\epsilon;t)=\Tr\{\hat{\rho}(\mathbf{\Omega},t)\hat{O}_{LM}(\mathbf{\Omega};\epsilon)\}$ . This provides a general route to quantum tomography for orientation dependent vibronic dynamics using MaxEnt. 
\subsection{Option 2: The Full Rovibronic Density Matrix}
The full rovibronic density operator in the symmetric top basis, $\ket{JMK}$, is
\begin{equation}
\hat{\rho}(t) = \sum_{ij}\sum_{J_iM_iK_i}\sum_{J'_jM'_jK'_j}\rho_{\omega_i\omega'_j}(i,j;t)\ket{i}\ket{J_iM_iK_i}\bra{J'_jM'_jK'_j}\bra{j}.
\end{equation}
Here $\omega\equiv \{J,M,K\}$, $J$ being the total angular momentum, $K$ its projection onto the molecule fixed axis and $M$ its projection onto the space fixed axis. In this basis the MADMs, $A^K_{QS}(i,j;t)$ can be written as
\begin{equation}
A^K_{QS}(i,j;t) = \sum_{\omega_i,\omega'_j}\rho_{\omega_i\omega'_j}(i,j;t)\left\langle{j\omega^{'}_j}\left|\hat{A}^{K}_{QS}\right|{i\omega_i}\right\rangle,
\end{equation}
where $\hat{A}^K_{QS}$ are previously defined Angular Momentum Coherence Operators. Their matrix elements\\ $\left\langle{j\omega^{'}_j}\left|\hat{A}^{K}_{QS}\right|{i\omega_i}\right\rangle$ in the symmetric top basis are analytically known~\cite{gregory2022}. With this, the measurement operator becomes,
\begin{equation}
\hat{O}_{LM}(\mathbf{\Omega};\epsilon)=\sum_{ij}\sum_{\omega_i \omega'_j}\sum_{KQS}C^{LM}_{KQS}(j,i;\epsilon)\left\langle{j\omega^{'}_j}\left|\hat{A}^{K}_{QS}\right|{i\omega_i}\right\rangle \ket{i}\ket{J_iM_iK_i}\bra{J'_jM'_jK'_j}\bra{j},
\end{equation}
and $\beta_{LM}(\epsilon;t)=\Tr\{\hat{\rho}(t)\hat{O}_{LM}(\mathbf{\Omega};\epsilon)\}$ gives the same result as Eq.~\ref{eq:data} for the measured quantity. Determining the operator $\hat{\rho}(t)$ is a larger problem than determining $\hat{\rho}(\mathbf{\Omega},t)$, since $\hat{\rho}(t)$ includes the $\ket{\Omega}$-off diagonal contributions explicitly excluded in Eq.~\ref{eq:LFDMop}. While this provides no new information on molecular frame dynamics, new lab frame quantities such as the Entropy of the molecule as a whole, or lab frame angular momenta become additionally accessible. Note, however, that $\hat{\rho}(\mathbf{\Omega},t)$ contains sufficient information to construct any molecular frame observable, and therefore the entanglement entropy of the electronic subsystem can still be constructed.    

\section{Procedure}

Here we apply MaxEnt to the NH$_3$ molecule resonantly excited to its two-electronic-state $\tilde{B}^1E''$ manifold by a femtosecond laser pulse; and probed by a subsequent, photoionizing laser pulse as it evolves in time. The LFDM was determined in~\cite{morrigan2023} with the help of spectroscopic selections rules. Specifically, the quantities $\beta_{00}(\epsilon,t)$, $\beta_{20}(\epsilon,t)$ and $\beta_{40}(\epsilon,t)$ from Eq.~\ref{eq:data} were directly measured for a fixed kinetic energy of the ejected electron ($\epsilon=0.2$~eV), as function of time after the excitation. The matrix elements of the measurement operator $C(i,j;\epsilon)$ are known from spectroscopic studies of this manifold, and symmetry selection rules were used to determine that only three, unique, MADMs $A^K_{QS}(i,j;t)$ were excited by the first pulse. Since three measurements were made at each time delay ($\beta_{00}$, $\beta_{20}$ and $\beta_{40}$) Eq.~\ref{eq:data} could be uniquely inverted to determine the MADMs and construct the LFDM (Eq.~\ref{eq:LFDMop}).\\
The unique reconstruction of a quantum system's density matrix in QIS, as detailed in \ref{entropy}, necessitates measurements in the form of expectation values derived from an informationally complete set of Hermitian operators. These operators are typically composed of the Pauli string operators ($\sigma_x$, $\sigma_y$, $\sigma_z$, $\sigma_i$). In contrast, here we use the operator in Eq.~\ref{eq:measop} with the MaxEnt density operator Eq.~\ref{eq:maxent} to directly determine the LFDM. At each time step, the process entails constructing the MaxEnt density matrix by parameterizing the measurement operators ($\hat{\beta}_{00}$, $\hat{\beta}_{20}$, and $\hat{\beta}_{40}$) with the Lagrange multipliers, as follows: 
\begin{equation}
    \hat{\rho}_{ME}(\Omega,\Vec{\lambda}(t)) = \frac{\exp\{-\lambda_{00}(t)\hat{\beta}_{00}(\Omega,t)-\lambda_{20}(t)\hat{\beta}_{20}(\Omega,t)-\lambda_{40}(t)\hat{\beta}_{40}(\Omega,t)\}}{Z(\lambda_{00}(t),\lambda_{20}(t),\lambda_{40}(t))} \label{eq:rho_beta}
\end{equation} 
The measurement operators vary with both the orientation angle ($\Omega$) and time, whereas the Lagrange multipliers are solely dependent on time. The partition function $Z(\lambda_{00}(t),\lambda_{20}(t),\lambda_{40}(t))$ ensures the normalization of the density operator and is obtained by taking the trace of the numerator in $\hat{\rho}_{ME}$ over all orientation angles at a specific time. The unknown Lagrange multipliers, at each time step, are subsequently determined using gradient descent optimization based on the three measurements acquired from experiments, represented as ($\beta_{00}$, $\beta_{20}$, and $\beta_{40}$). The associated cost function is defined as: 
\begin{equation}
    \textrm{Cost} = (\sum_{LM} Tr\{\hat{\rho}_{ME}(\Omega,\Vec{\lambda}(t))\hat{\beta}_{LM}\} - \beta_{LM})^2
\end{equation}
Essentially, the cost function quantifies the mean-squared error (MSE) between the actual and computed expectation values of the measurement operators, derived from the reconstructed MaxEnt density operator. The optimization iterates until the cost falls below a predetermined threshold, resulting in obtaining the converged set of Lagrange multipliers. Upon determining the Lagrange multipliers, Eq. \ref{eq:rho_beta} represents the resulting LFDM.

\section{Results}
\subsection{Quantum Tomography by the Maximum Entropy Approach}
\begin{figure}
    \centering
    \includegraphics[scale=0.45]{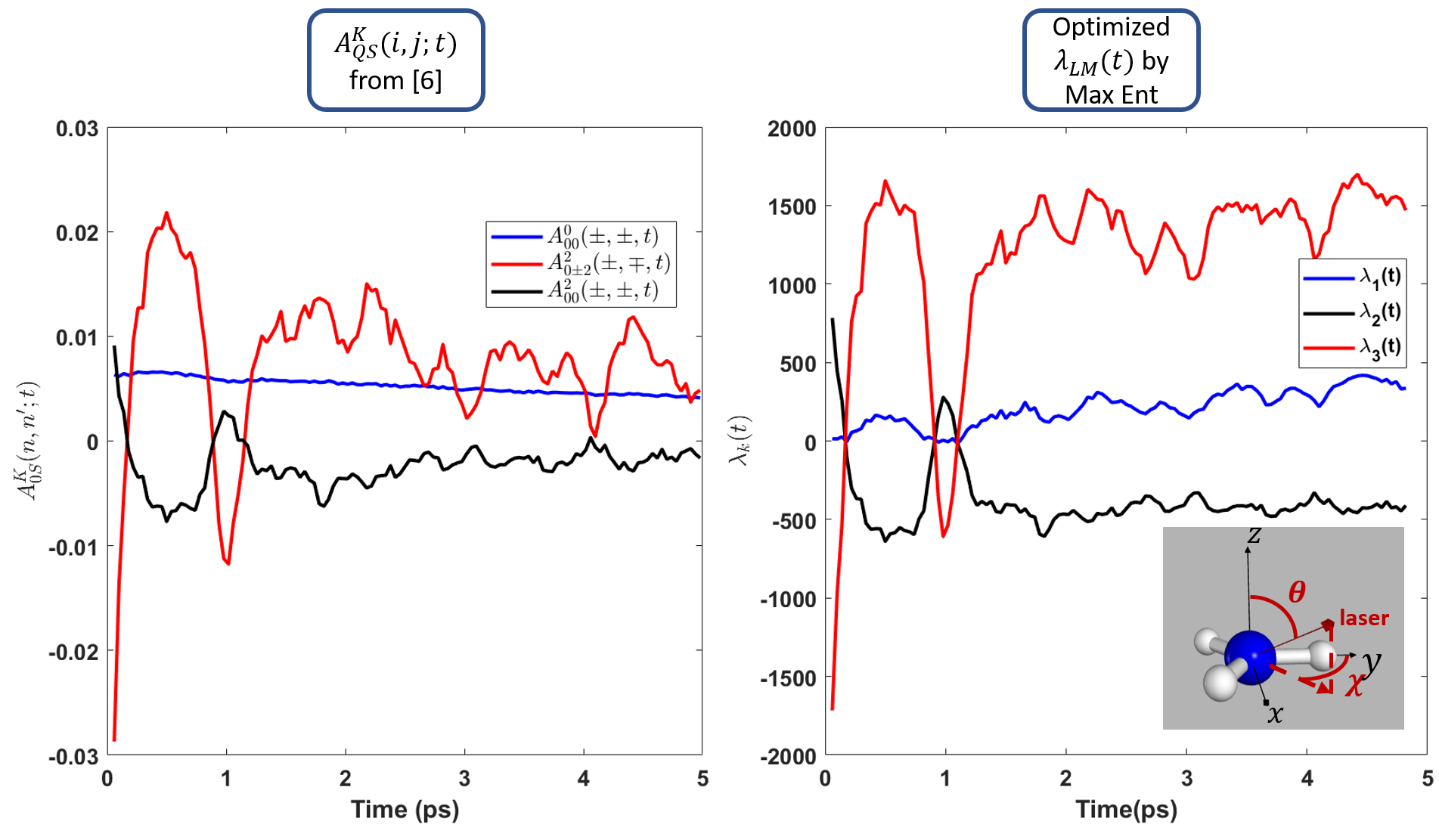}
    \caption{A comparison of the optimized $\lambda_k(t)$ parameters from the MaxEnt procedure and the unique MADMs $A^K_{QS}(i,j;t)$ determined in Morrigan et al.~\cite{morrigan2023}. This comparison reveals that the MaxEnt procedure automatically determines the unique set of MADMs that determine the excited state molecular dynamics. The inset at the bottom right shows the geometry of NH$_3$ in the $\tilde{B}^1E''$ manifold with the relevant orientation angles $\mathbf{\Omega}=\{\theta,\chi\}$ with respect to polarization of the excitation laser pulse.}
    \label{fig:lmbdas}
\end{figure}

In Fig.~\ref{fig:lmbdas} the optimized Lagrange Multipliers resulting from the MaxEnt procedure alongside the three unique MADMs determined in~\cite{morrigan2023}. Comparison of these parameters provides an immediate interpretation of the $\lambda_k(t)$ in Eq.~\ref{eq:maxent}: they are the unique set of MADMs $A^K_{QS}(i,j;t)$ that determine the excited state dynamics of the molecule. We reiterate that to perform the tomography in~\cite{morrigan2023}, the number of unique non-zero MADMs and their labels needed to be determined before hand. The MaxEnt procedure does not label the MADMs; that is, the values of $K, Q, S$ and the pair of electronic states $\ket{i}$, $\ket{j}$ associated with each $\lambda_i(t)$ are not known. However, it automatically determines the correct set of MADMs to within a scale factor. In this particular case of NH$_3$ the three $\lambda_k(t)$ correspond to: (i) the coherence - $A^2_{0\pm2}(\pm,\mp;t)$ - between the pair of electronic states here labelled $\ket{+}$, $\ket{-}$; (ii) the average alignment of the 3-fold symmetry axis of NH$_3$ -  $A^2_{00}(\pm,\pm;t)$ - in each electronic state; and (iii) the orientation integrated population -  $A^0_{00}(\pm,\pm;t)$ - of each electronic state. Additional studies are required to determine whether this correspondence between the $\lambda_k(t)$ and coherence, molecular alignment and population is general; but the results presented here provide a first interpretation of these Lagrange multipliers.

\begin{figure}
    \centering
    \includegraphics[scale=0.50]{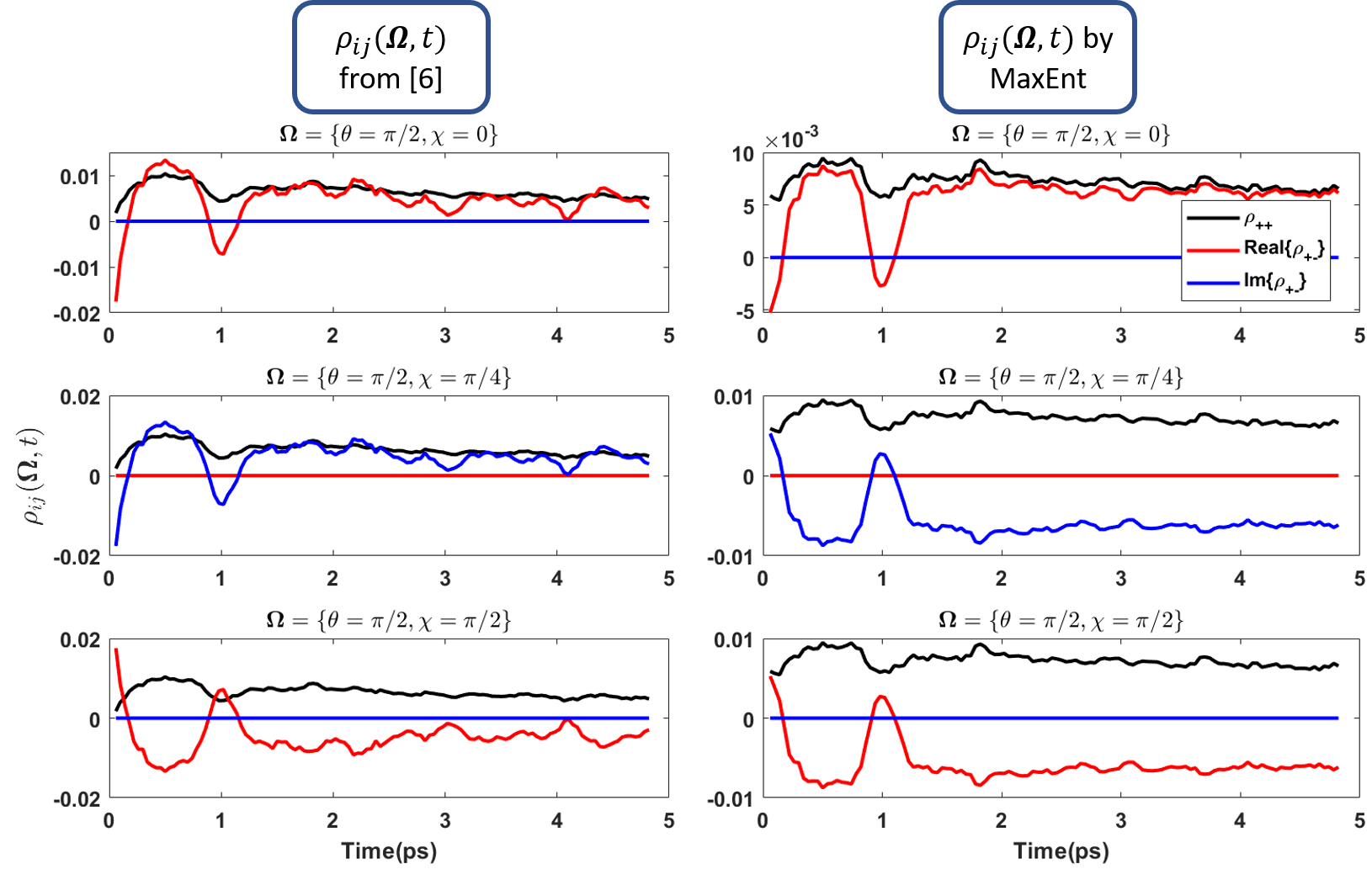}
    \caption{A comparison of the LFDM determined by MaxEnt to that determined in~\cite{morrigan2023} for different orientations of the molecular frame relative to the polarization of the excitation laser pulse (see Fig.~\ref{fig:lmbdas} inset).}
    \label{fig:LFDM}
\end{figure}

In Fig.~\ref{fig:LFDM} we compare the LFDM determined by MaxEnt to that determined in~\cite{morrigan2023} at three selected orientations. A number of qualitative features are in agreement: (i) the time evolving populations are identical at each of the three orientations; (ii) the electronic coherence varies rapidly in the first 1.5~ps after excitation, after which it remains relatively stable with small fluctuations; (iii) the phase of the coherence changes with orientation in the same manner; and (iv) the real or imaginary parts of the coherence vary with significantly lager amplitude than do the populations at all orientations. These are the primary determining features of the electronic and rotational dynamics of the molecule as discussed in detail below. The noteworthy quantitative differences between the two results are an overall offset in the populations, as well as a scale factor and greater stability in the coherence. These are primarily due to the fact that the $\tilde{B}^1E''$ manifold of NH$_3$ is known to be dissociative, which is reflected in the slow decay of the population and coherence from~\cite{morrigan2023}. The MaxEnt procedure strictly requires that $\Tr\{\rho(\mathbf{\Omega},t)\}=1$, thus compensating for the slow decay of experimental signal. As an advantage, this allows the construction of the entanglement entropy of the excited electron, which was not possible with the result from~\cite{morrigan2023}, as we show below.
\subsection{Molecular Dynamics}
\begin{figure}
    \centering
    \includegraphics[scale=0.45]{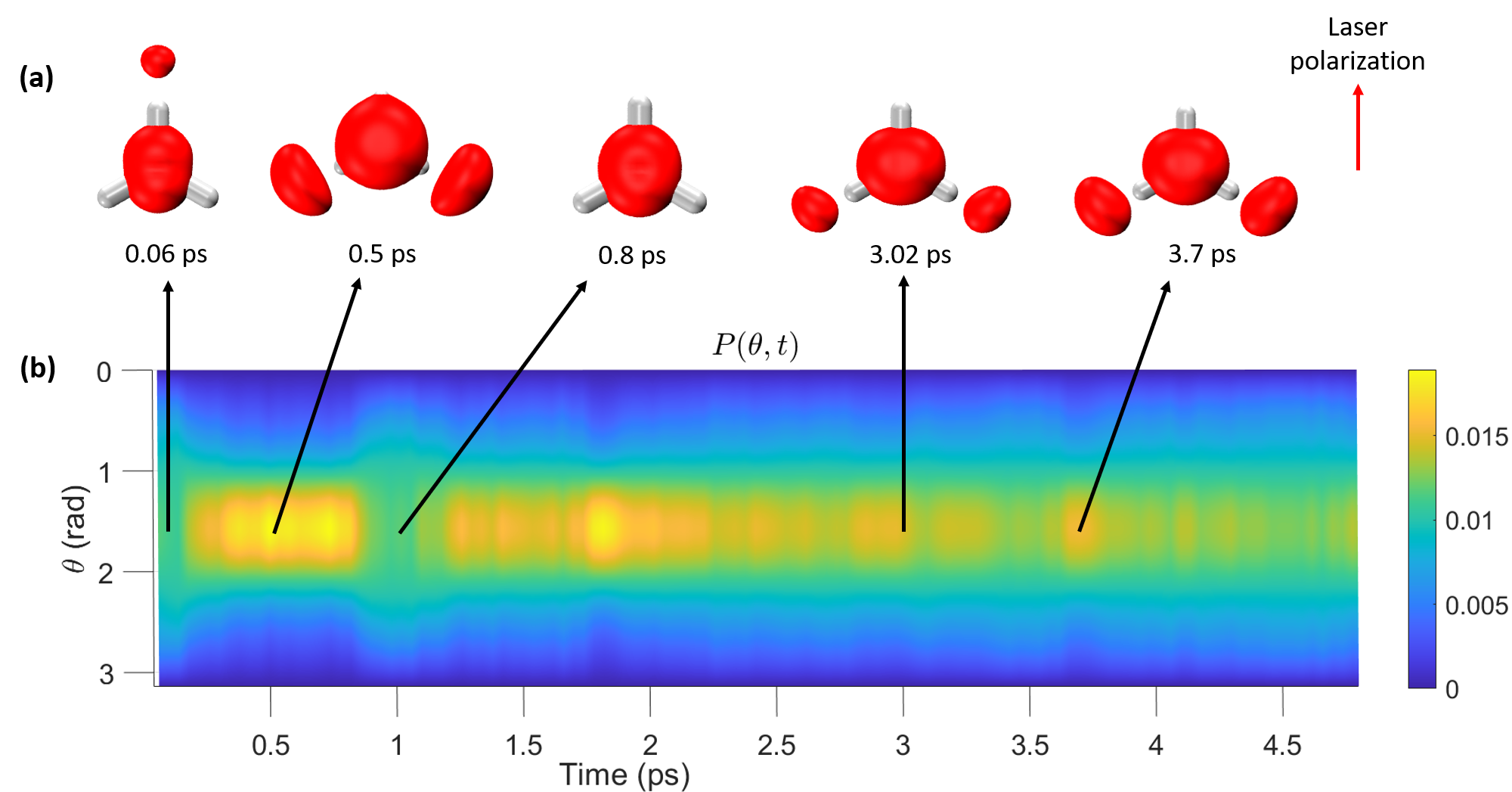}
    \caption{The time dependent (a) one electron attachment density at orientation $\mathbf{\Omega}=\{\theta = \pi/2, \chi= 0$\}  and (b) 
 the molecular axis distribution constructed from the MaxEnt LFDM. The former reveals the electronic motion in the molecular frame as charge migration and the latter tracks the rotational motion of the molecule. }
    \label{fig:dynamics}
\end{figure}

\begin{figure}
    % \centering
    \includegraphics[scale=0.50]{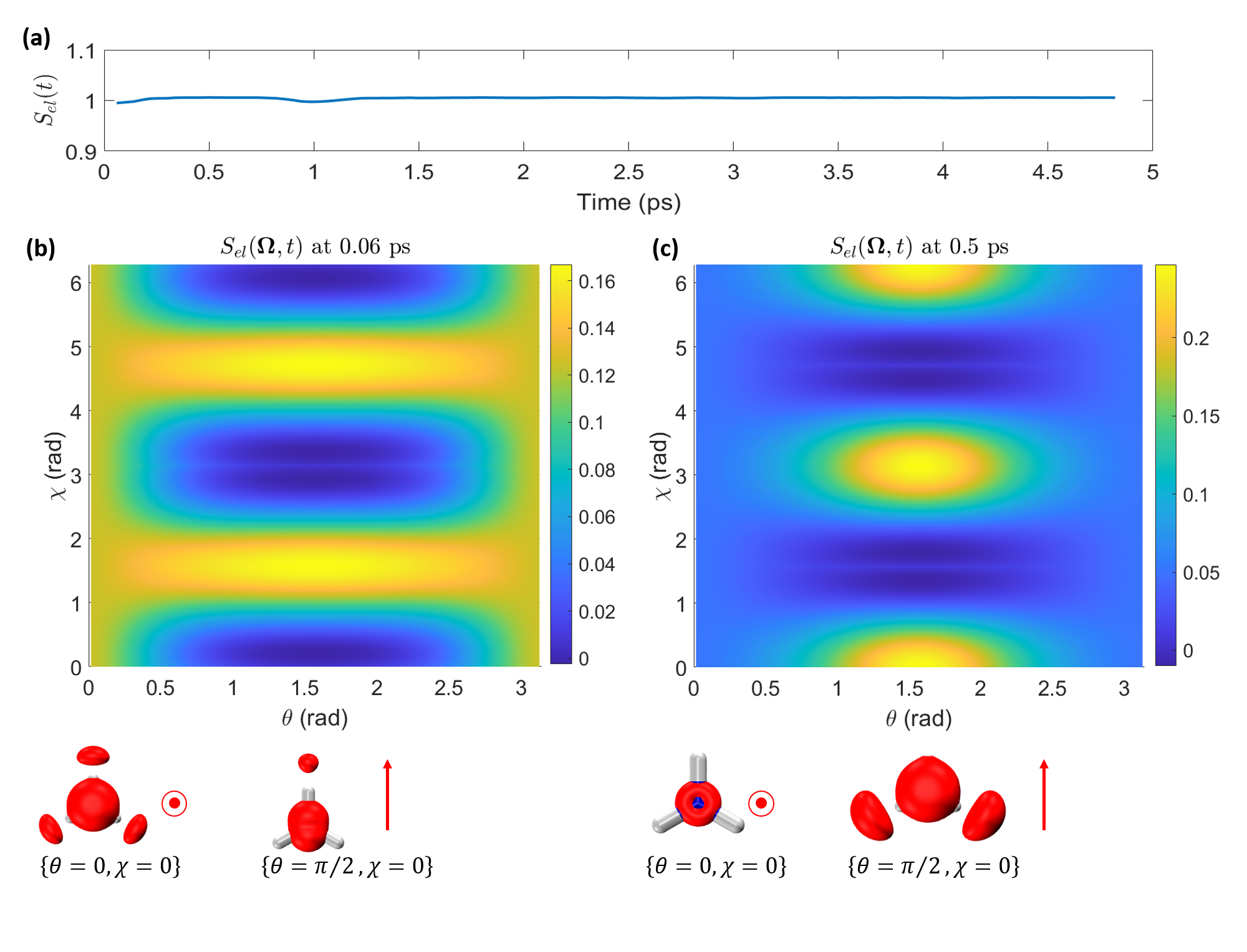}
    \caption{(a) The entanglement entropy of the electron as a function of time constructed using the MaxEnt LFDM, indicating that the electronic state is in a maximally mixed or entangled state. The orientation dependent entropy of the electron at (a) 0.06 ps and (b) 0.5 ps after the pump pulse, along with one-electron attachment densities at $\mathbf{\Omega}=\{\theta = 0, \chi= 0$\} and $\mathbf{\Omega}=\{\theta = \pi/2, \chi= 0$\} constructed using the MaxEnt LFDM. These maps show the effect of coupling between the electron and rotating nuclei, with high entropy regions indicating a highly mixed electronic-nuclear state.}
    \label{fig:entropy1}
\end{figure}

With the LFDM determined, we can construct the electron density in position space, $P(\mathbf{r},\mathbf{\Omega},t) =  \\\sum_{ij}\rho_{ij}(\mathbf{\Omega},t)\phi_{i}(\mathbf{r})\phi_{j}(\mathbf{r})$, with $\mathbf{r}$ the position vector of the electrons in the molecular frame and $\phi_{i}(\mathbf{r})$ the wavefunction associated with the electronic state $\ket{i}$. In Fig~\ref{fig:dynamics}(a) we show the one-electron attachment density~\cite{gregory2021,morrigan2023} at selected time delays, for the NH$_3$ molecule with polarization of the excitation pulse along an NH bond. The attachment density quantifies the accumulation of density in the molecular frame~\cite{morrigan2023}. As in~\cite{morrigan2023} we observe charge migration across the molecule with a period of about 1~ps at this orientation. In Fig~~\ref{fig:dynamics}(b) we show the molecular axis distribution $P(\theta,t)=\sum_{i}\rho_{ii}(\theta,t)$ as a function of time, $\theta$ being the relative angle of the molecular 3-fold symmetry axis and the polarization of the excitation pulse (see Fig.~\ref{fig:lmbdas} inset). In agreement with~\cite{morrigan2023} we observe rapid rotation of the 3-fold axis in the first 1.5~ps, synchronized with the charge migration in the molecular frame, followed by small fluctuations around $\theta = \pi/2$ for the remainder of the time window. In~\cite{morrigan2023} this is attributed to ro-electronic coupling due to a molecular Coriolis effect.   
\subsection{Entropy of the Electron: Electron-Nuclear Coupling}
Here we can additionally provide direct evidence of ro-electroinc coupling by constructing the entanglement entropy of the electronic subsystem, $S_{el}(t) = -\Tr{\hat{\rho}(\Omega,t)\log{\hat{\rho}(\Omega,t)}}$, shown in Fig.~\ref{fig:entropy1}(a). Evidently, $S_{el}(t) \approx 1$ for the entire time window. To interpret this we can consider the spectroscopic model of the $B^{ 1}E''$ state developed by Western et al.~\cite{ashfold1998,Ashfold1987}. To accurately model the absorption spectrum NH$_3$ is treated as a rotating, planar, symmetric top coupled to a single excited electron with states $\ket{+}$ and $\ket{-}$. In this bi-partite model, the measured $S_{el}$ suggests that the excited electron is in a maximally mixed, or maximally entangled state. Prior to excitation, however, NH$_3$ is in the ground electronic state, so $S_{el} = 0$.The excited state is prepared coherently by the excitation laser pulse, implying that a measurement of $S_{el}\approx 1$ in fact suggests maximal entanglement of the excited electron with the rotating molecular core. To determine the characteristics of this ro-electronic coupling, we can finally construct an orientation dependent entropy of the excited electron, $S_{el}(\Omega,t) = -\Tr{\hat{\rho}(\Omega,t)\log{\hat{\rho}(\Omega,t)}}$, without tracing over orientation angles, as shown in Fig.~\ref{fig:entropy1}(b) and (c). This provides a map of the degree of ro-electronic mixing as a function of orientation and time, with the orientation averaged entropy remaining constant (Fig.~\ref{fig:entropy1}(a)). We observe that orientations of maximum entropy - where the electronic state is highly mixed, or correlated with molecular rotation -  change considerably within the first 0.5~ps. Furthermore, at high entropy orientations the electronic density is more spread out over the molecule as shown in the insets of Fig.~\ref{fig:entropy1}(b) and (c); at 0.06~ps for $\{\theta = 0, \chi = 0\}$ and at 0.5~ps for $\{\theta = \pi/2, \chi = 0\}$. On the other hand , for orientations with minimum entropy, the electronic density is more compact; for $\{\theta = \pi/2, \chi = 0\}$ at 0.06~ps it is polarized along the NH bond, parallel to the laser polarization and for $\{\theta = 0, \chi = 0\}$ at 0.5~ps is localized at the N atom. By about 1.5~ps $S_{el}(\Omega,t)$ reverts to the orientation dependence shown in Fig.~\ref{fig:entropy1}(a), and finally stabilizes to the dependence in Fig.~\ref{fig:entropy1}(b) in concert with stabilization of the rotational dynamics ($P(\theta,t)$ in Fig.~\ref{fig:dynamics}). In sum, this provides a complete picture of the electronic dynamics in the molecular frame and its coupling to nuclear motion: the evolution of the electronic density at any orientation is driven by coupling to the rotational subsystem as quantified by $S_{el}(\Omega,t)$. When the electronic state is highly mixed with that of the rotating molecule the charge density spreads over the entire molecular frame and it localizes when the electronic state becomes approximately separable. \\Alternatively, $S_{el}(\Omega,t)$ provides a map of the flow of quantum information in the space of molecular orientations. In this case, Fig.~\ref{fig:entropy1}(a) indicates that information does not dissipate from the electronic subsystem within the time window of the experiment, even though its state is mixed with that of the rotating nuclei. The ro-electronic coupling only causes the flow of information between different molecular orientations. Quantum information dynamics in molecules is a relatively new topic of research~\cite{lewis2019,zhang2022,liu2023,zhang2024}, and we show here the MaxEnt approach can provide access to metrics of quantum information in isolated molecules.

\section{Conclusion and Outlook}

Our work provides a framework to apply quantum tomography methods developed in QIS to the ultrafast dynamics of isolated molecules. In doing so it provides the tools necessary to image electronic dynamics in the molecular frame, a general goal in attosecond and molecular physics, providing insight into the coupling of these dynamics to nuclei as well as the ability to monitor the correlation of the electrons and the nuclei. Here, we have demonstrated these capabilities by applying the MaxEnt approach to quantum tomography in the NH$_3$ molecule resonantly excited to its $\tilde{B}^1E''$ manifold of states. While being consistent with previous results, this approach also provides the ability to construct the entropy of the electron in the molecular frame revealing the consequences of coupling between the excited electron and the rotating nuclei. These results potentially open new areas of inquiry addressing the flow of quantum information in molecules. Here, we observe that electronic-rotational coupling redistributes information in the space of orientations as function of time, without causing dissipation. An immediate application of this approach would be to study the dynamics of quantum information in the nearby, Jahn-Teller active, '$\tilde{C}$' manifold~\cite{allen1991} of NH$_3$ activating coupling between electronic motion and vibration, as well as between vibrational modes. Recent theoretical studies suggest that such interactions could lead to a scrambling of quantum information~\cite{zhang2022,zhang2024}. Additionally we note that any approach to quantum tomography requires determination of the observable operator $\hat{O}_{LM}(\mathbf{\Omega};\epsilon)$ representing the probe of molecular dynamics. This has been determined for one photon ionization of only a handful of excited molecular states by high resolution spectroscopy~\cite{reid1992,motoki2002,lebech2003,cherepkov2005,teramoto2007,hockett20071,hockett20072,hockett2009,marceau2017}. However, recently developed methods in attosecond physics that determine the state of an ejected photoelectron~\cite{Laurent2012,hockett2017AngleresolvedRABBITTTheory,vrakking2021,laurell2022,laurell2023} can potentially be applicable here. As such, our work presents numerous opportunities for further investigation in the fields of ultrafast molecular dynamics and QIS.

\section*{Acknowledgements}
We would like to thank Raphy D. Levine and Manas Sajjan for many useful
discussions.  S.K. would also like to acknowledge funding from the U.S. Department of Energy (DOE) (Office of Basic Energy Sciences), under Award No. DE-SC0019215, and the National Science Foundation under Award No. 1955907.

\bibliographystyle{unsrt}
% \bibliography{refs, NH3, MatEtheory}
\bibliography{refs}

% Testing bib stuff...
% This might work, but not on Overleaf!
% See https://tex.stackexchange.com/a/355331
% \input{si.bbl}

\end{document}